\documentclass[manuscript]{aastex}
\usepackage{amssymb}

\usepackage{amsmath}

\slugcomment{Not to appear in Nonlearned J., 45.}

\shorttitle{Fine Magnetic Characteristics of a Light Bridge}
\shortauthors{S. Liu et al.}

\begin{document}


\title{A Full-disk Image Standardization of the Chromosphere Observation at Huairou Solar Observing Station}

\author{S. Liu\altaffilmark{1,2}}

\affil{
$^{1}$Key Laboratory of Solar Activity, National Astronomical Observatories, Chinese Academy of Science, Beijing, 100101, China\\
$^{2}$School of Astronomy and Space Sciences, University of Chinese Academy of Sciences, Beijing, 100101, China}

\email{lius@nao.cas.cn}

\altaffiltext{1}{key Laboratory of Solar Activity}

\begin{abstract}
Observations of local features in the solar chromosphere began in 1992 at Huairou Solar Observing Station, while the full-disk chromosphere observations were carried out since 2000.
In order to facilitate researchers to use full-disk chromosphere observation, algorithms have been developed to standardize the full-disk images.
The algorithms include the determination of the center of the image and size standardization, geometric correction and intensity normalization. The solar limb of each image is determined from a histogram analysis of its intensity distribution. The center and radius are then calculated and the image is corrected for geometric distortions. Images are re-scaled to have a fixed radius of 500 pixels and centered within the 1024$\times$1024 frame. Finally, large-scale variations in intensity, such as limb-darkening, are removed using a median filter. This paper provides a detailed description of these algorithms, and a summary of the properties of these chromosheric full-disk observations to be used for further scientific investigations.
\end{abstract}
\section{Introduction} \label{sec:intro}
There are varioius solar activities, such as flare, filament eruption and corona mass ejection (CME), in the atmosphere of Sun. Moreover, different phenomena are often interrelated. For example, radiation enhancemen of remote chromospheret related solar flare (\citeauthor{1982SoPh...77..263T}, \citeyear{1982SoPh...77..263T}; \citeauthor{2017SoPh..292...72K}, \citeyear{2017SoPh..292...72K});
sequential chromospheric brightenings and sympathetic eruptions among flares (\citeauthor{2005ApJ...630.1160B},\citeyear{2005ApJ...630.1160B}; \citeauthor{2017SoPh..292...72K},\citeyear{2017SoPh..292...72K}); small-scale activities and associated flares (\citeauthor{1983PASJ...35..285K}, \citeyear{1983PASJ...35..285K}; \citeauthor{2007A&A...472..967C}, \citeyear{2007A&A...472..967C}; \citeauthor{2012ApJ...760...31K}, \citeyear{2012ApJ...760...31K}; \citeauthor{2017NatAs...1E..85W}, \citeyear{2017NatAs...1E..85W}; \citeauthor{2018SoPh..293..167L}, \citeyear{2018SoPh..293..167L}). The above statements give an impression that large-scale structures should be responsible for these interrelated phenomena, as a result systematic studies of these large-scale events are essential to understand the topological structures and  corresponding eruptive events. Consequently, what follows is the need to obtain high-quality full-disk observations. Recently, significant progress has been made in the detection of filament and statistics studies of filaments and their eruptions (\citeauthor{2011SoPh..272..101Y}, \citeyear{2011SoPh..272..101Y}; \citeauthor{2013SoPh..286..385H}, \citeyear{2013SoPh..286..385H}; \citeauthor{2016SoPh..291.1115T}, \citeyear{2016SoPh..291.1115T}), for which most of studies are based on full-disk chromosphere observations
with high-quality.

Due to the importance of full-disk observations for understanding the complex solar activities, full-disk chromosphere observations are carried out by most parts of ground solar observatories. Such as  Kodaikanal Solar Observatory (India), Meudon Observatory (France), National Solar Observatory–Sacramento Peak and Big Bear Solar Observatory (USA), Kislovodsk Observatory (Russia), Kanzelh\"{o}ehe Observatory (Austria), Yunnan Astronomical Observatory and Huairou Solar observing Station (China). Additionally, Global High Resolution H-alpha Network integrates solar chromosphere H$\alpha$ observations around the world to make the data more convenient to use.
However, there was hardly internationally accepted standard for standardizing H$\alpha$ full-disk images, while the standardization process is needed to facilitate the use of observed data.
\citeauthor{2003SoPh..214...89Z} (\citeyear{2003SoPh..214...89Z}) developed techniques, which corrected non-uniformity of the disk shape and intensity, to process the H$\alpha$ and Ca K line full-disk images taken at the Meudon Observatory into a standardised form. For each full-disk image the fitting of limb, removal of geometrical distortion, centre position and size standardization and intensity
normalization are included in the methods used. \citeauthor{2013SoPh..286..385H} (\citeyear{2013SoPh..286..385H}) developed an advanced method to automatically detect and trace solar filaments in H$\alpha$ full-disk images obtained by Mauna Loa Solar Observatory (MLSO), in which the basic pre-processings/standardization were carried out on orignal data such as limb-darkening removal, solar disk extraction and filter enhancement. Providing homogeneous sets of full-disk images can benefit the completion of tasks such as feature recognition and large data batch processing (\citeauthor{2005SoPh..227...61F}, \citeyear{2005SoPh..227...61F}; \citeauthor{2005SoPh..228..137Z}, \citeyear{2005SoPh..228..137Z}; \citeauthor{2005SoPh..228..149L}, \citeyear{2005SoPh..228..149L}; \citeauthor{2010ApJ...717..973W}, \citeyear{2010ApJ...717..973W}; \citeauthor{2012SoPh..275...79M}, \citeyear{2012SoPh..275...79M}; \citeauthor{2013SoPh..286..385H}, \citeyear{2013SoPh..286..385H}; \citeauthor{2015ApJS..221...33H}, \citeyear{2015ApJS..221...33H}; \citeauthor{2015SoPh..290.1963L}, \citeyear{2015SoPh..290.1963L}, \citeauthor{atoum2017}, \citeyear{atoum2017}).

In this paper, the algorithms are introduced that have been developed to standardize full-disk H$\alpha$ chromosphere observations obtained by HSOS, to maximize the exploration of these chromosphere observations. The description of chromosphere observations is arranged in Section~\ref{S-Obser and Data};
then, the details of techniques and examples about standardization algorithms are given in Section~\ref{S-Algorithms}; finally, section~\ref{S-Conl} presents the discussions and conclusions.

\section{OBSERVATIONS}\label{S-Obser and Data}
\begin{figure}
   \centerline{\includegraphics[width=0.8\textwidth,clip=]{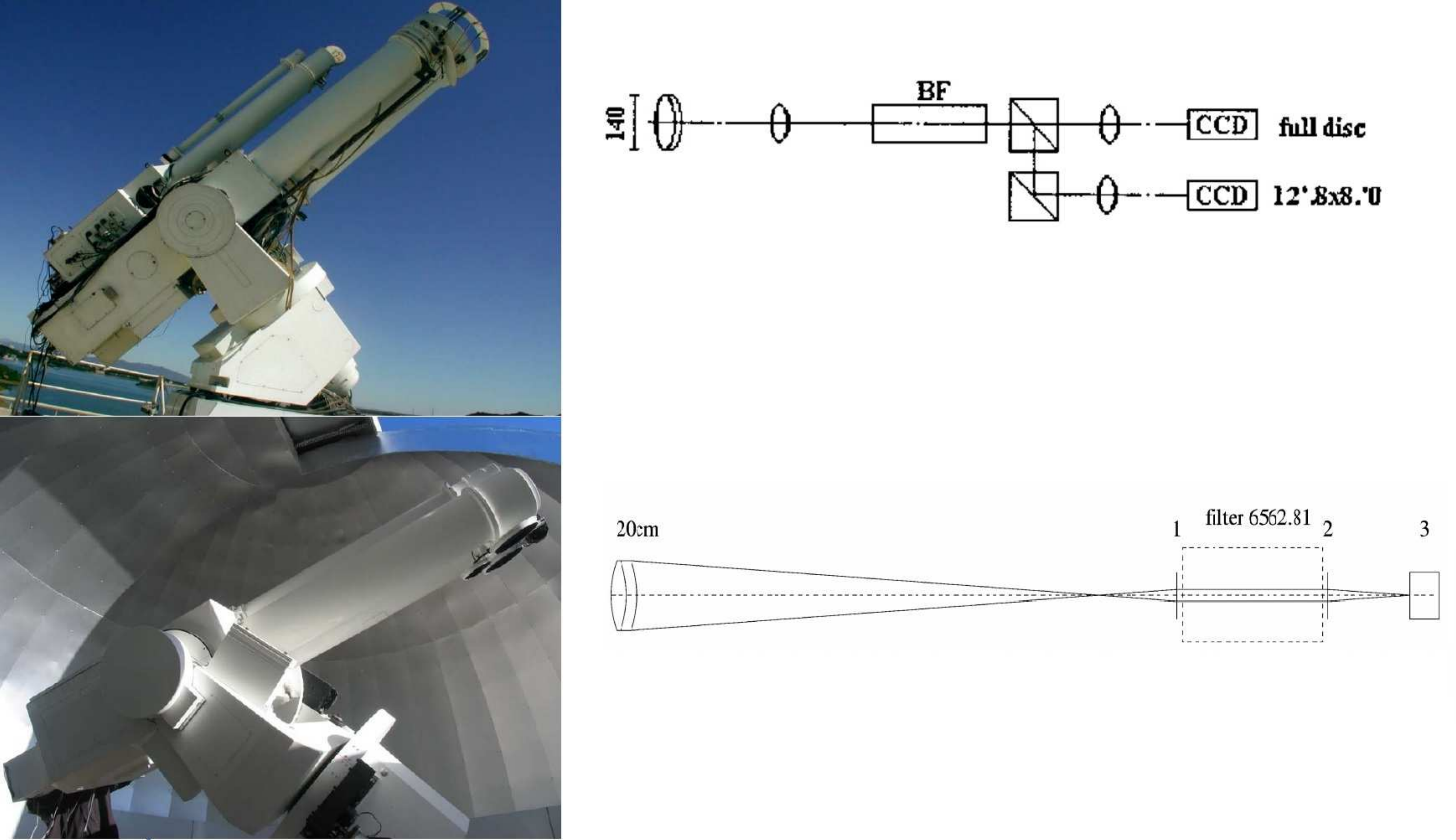}}
   \caption{Upper: the SMCT (left) and the optical scheme of 14-cm full disk  H$\alpha$ telescope (right). Bottome: the SMAT (left) and the optical scheme of 20-cm full disk  H$\alpha$ telescope (right). } \label{telescope}
\end{figure}

Huairou Solar observing Station (HSOS: \href{http://sun.bao.ac.cn}{http://sun.bao.ac.cn}), which is located at the north shore of Huairou reservior (Latitude:41.3, Longitude:116.6), is a main solar observatory in the world. HSOS was built in 1984, and can provide many kinds of solar data, such as active region vector magnetic field and Doppler velocity in photosphere, active region longitudinal magnetic field and Doppler velocity in chromosphere, full-disk magnetic field in the photosphere, local and full-disk H$\alpha$ and Ca II (3933.8\AA) image in chromosphere. So far, the observations with three solar cycles have been carried out, it has brought abundant of historical observation data to solar physics research. The routine local chromosphere observations started in 1992, while the full-disk chromosphere H$\alpha$ (6562.8\AA) observations were carried out since 2000, and it operates up to now. Before 2006, full-disk chromosphere H$\alpha$ images were observed by a 14-cm telescope equipped with a tunable H$\alpha$ filter filer (-32\AA~ +32\AA~with bandwidth 0.5\AA) and a KODAK MegaPlus 4.2i Camera for full-disc H$\alpha$ monochromatic image. Full-disk images cover a 34$^{'}$~$\times$~34$^{'}$ field-of-view and their image size is 2106x2044 pixels. This 14-cm telescope is one of optical telescope/system of the Solar Multi-Channel Telescope (SMCT; \citeauthor{1986AcASn..27..173A}, \citeyear{1986AcASn..27..173A}; \citeauthor{1997SoPh..173..207D}, \citeyear{1997SoPh..173..207D}) at HSOS, sampling frequency of full-disk H$\alpha$ images is about a few images a day, and the images were stored on disk in DAT format. After 2006, a new 20-cm telescople, which belongs to the Solar Magnetism and Activity Telescope (SMAT; \citeauthor{2007ChJAA...7..281Z}, \citeyear{2007ChJAA...7..281Z}), was built to observe Full-disk H$\alpha$ images. A birefringent filter for the H$\alpha$ observations is centered at 6562.8\AA~and its bandpass is 0.25\AA. The center wavelength of the filter can be tuned within $\pm$2\AA~from the H$\alpha$ line center.
A CCD camera (Kodak KAF-4202) is used for the measurement of full disk H$\alpha$ filtergrams. The image size of the telescope is 9 mm $\times$ 9 mm, and the size of the CCD is 2029 $\times$ 2044 pixels (now 2712 $\times$ 2712 pixels). The spatial resolution of full disk H$\alpha$ filtergrams is about 2 arecsec and series of images can be observed continuously. For routine observation of 20-cm telescope, the time resolution of full-disk H$\alpha$ images is between 1-5 minutes based on scientific objectives, and the images are stored on disk as standard Flexible Image Transport System (FITS: http://fits.gsfc.nasa.gov/) format in real time. For full-disk observations, they were made to record activity on the solar disk, however, if there is flare the local region is outlined that can be observed with high cadence and reasonable exposure time. Figure \ref{telescope} shows the SMCT/SMAT at HSOS and the optical scheme of 14/20-cm full disk  H$\alpha$ telescope, respectively.
\section{ALGORITHMS AND RESULTS}
\label{S-Algorithms}
For the full-disk H$\alpha$ at HSOS, there exist inconveniences for scientific research due to the image variations originated from data format, CCD size, optical modifications and data acquisition system and multi-observer operations. Hence, some procedures and algorithms are developed to standardize these observations data, addtionally some necessary pre-processing should be included in the algorithms. The following sections give the details of techniques and examples about standardization algorithms.

\begin{figure}
   \centerline{\includegraphics[width=1\textwidth,clip=]{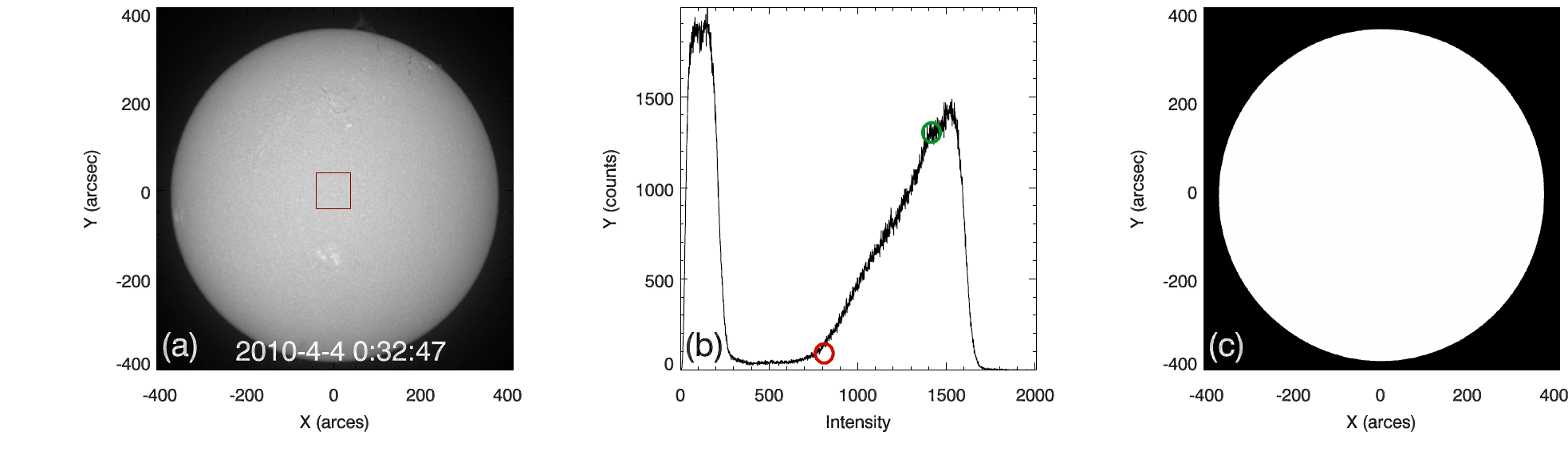}}
   \caption{(a): full-disk  H$\alpha$ image, the red square region is calculated for solar intensity. (b): histogram of the whole image observed, the green circle represent the solar intensity and the red circle indicates the value of threshold estimated. (c): processed full-disk image with 1 in solar disk and 0 background.} \label{threshold}
\end{figure}
\begin{figure}
   \centerline{\includegraphics[width=0.8\textwidth,clip=]{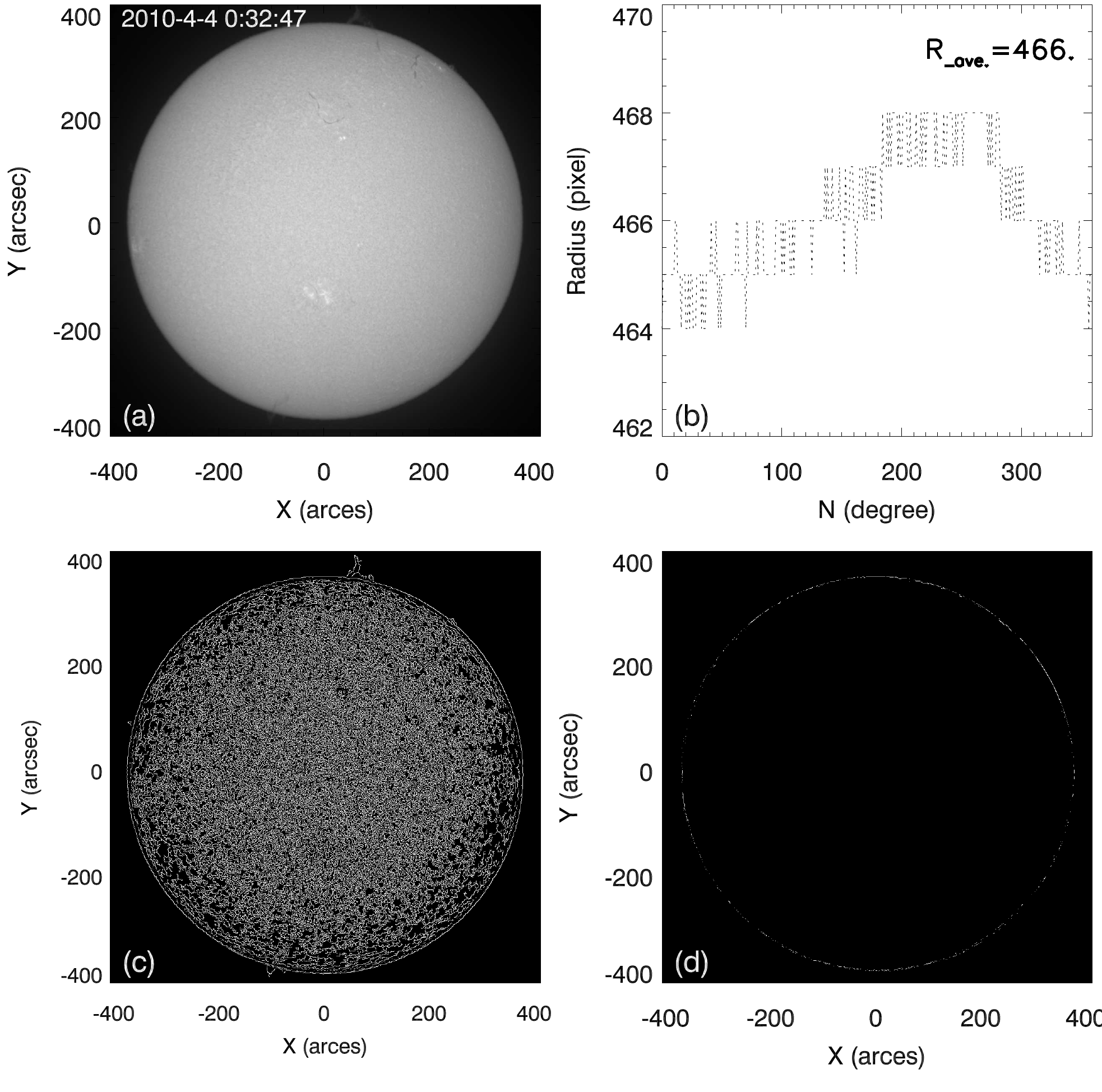}}
   \caption{(a): The shifted full-disk image based on the initial center from centre-of-gravity method. (b): The distributions of multi-radii calculated by center-of-gravity method, the average of radius are labeled with the value of 466 pixels. (c): The edges detected by canny method using shifted image (a) as input data.(d): The corresponding higher-presion limb of solar disk obtained by limited the average of multi-radius and edges detected by canny method.} \label{cenrad}
\end{figure}
\subsection{Limb Threshold}
The intensity value of solar limb in individual full-disk image is very important for later data processing, such as the generation of mask data. Here the estimation of limb threshold is based on histogram analysis of full-disk image intensities (\citeauthor{2003SoPh..214...89Z}, \citeyear{2003SoPh..214...89Z}), the specific process is as follows: Firstly, the mean intensity of a central region the 10\% of the size of the whole image (Figure \ref{threshold}a) is obtained and noted as solar intensity, which can approximately segment the full-disk from the background. Then, the threshold of solar limb can be estimated by starting from solar intensity (Figure \ref{threshold}b, green circle) in the histogram plot and moving along the plot to direction of lower intensity, and stop at the position where the number of pixels is less than 8\% (this value should be tested, for these data is 8\%) the total number of pixel contained in the whole image and the number of pixel is starting to rise, the intensity of this stopped position (Figure \ref{threshold}, red circel) is regared as the reasonable threshold. Lastly, the pixels with value higher than the threshold are replaced by 1, and other pixels are filled 0. Figure \ref{threshold} illustrates the these process, (a): it shows a full-disk  H$\alpha$ image observed at HSOS, in the approximate center of full-disk image the red square is cut and calculated for solar intensity (Idisk). (b): it shows histogram of the value of the whole image, the red circle indicates the value of threshold estimated. (c): it gives processed full-disk image based on histogram analysis, and it is filled with 1 in solar disk and 0 background.

\subsection{The Center and Radius Calculation for full-disk image}
\label{sub-cenrad}

First, the center of gravity method is used here to compute the initial coordinates $X_c$ and $Y_c$ of the full-disk processed image (Figure \ref{threshold}c) center, that is:
\begin{equation}
X_c = \sum_{i=1}^{N}X_iI_i/\sum_{i=1}^{N}I_i
\end{equation}
\begin{equation}
Y_c = \sum_{i=1}^{N}Y_iI_i/\sum_{i=1}^{N}I_i,
\end{equation}
where $N$ is the total number of pixels in the image, $X$ and $Y$ are the pixel coordinates, and $I$ is the value (0 or 1) of that pixel.
Second, starting from the position of center of disk move radially toward outside of disk and find the boundary that divides solar disk (pixel with value of 1) from background (pixel with value of 0), so the initial multi-radii (here 360 groups meaning 360$^{\circ}$ set) can be calculated (the radius is the distance between center and boundary of disk), consequently the radius can be obtained by averaging these initial multi-radii (r= $(\sum_{n=1}^Nr_i)/N$, here N=360).
Third, the original image is re-centered based on the initial estimate of its center. The canny edge detector (\citeauthor{canny}, \citeyear{canny}) is then applied to detect the limb of the image using: 1. Gaussian filter with a 5x5 kernel ($\sigma$= 0.6) is used to smooth the image. 2. Nonmaxima suppression is used, where an edge point is defined to be a point whose gradient magnitude is locally maximum in the gradient direction. 3. The default values of high and low threshold is set to 0.8 and 0.4, respectively. 4. Hysteresis is applied to eliminate gaps in the processed image. The expected limb is determinded by limiting the distance between points found by canny operator and the shifted center of disk is equal to radius (averaging from multi-radii, here distance - radius $\le$ 1 pixel). At last, the expected limb points are used by circlefit.pro (Solar Software-SSW) method to obtain the high-precision center and radius (sigle-value radius fitted with high-precision) of full-disk image. Fiugre \ref{cenrad} shows the above processes briefly: (a) it shows the shifted full-disk image based on the initial center. (b) it gives the distributions of multi-radii calculated by center-of-gravity method, with the average of radius 466 pixels, and here the number of initial multi-radii is 360 (meaning 360$^{\circ}$ divided by 1$^{\circ}$). (c) it shows the edges detected by canny method using shifted image as input. (d) it indicates the corresponding higher-presion limb of solar disk obtained by the results of comprehensive restrictions between average of multi-radius and edges detected by canny method.


\subsection{Field-of-view and Circle Correction Processing}
\begin{figure}
   \centerline{\includegraphics[width=1\textwidth,clip=]{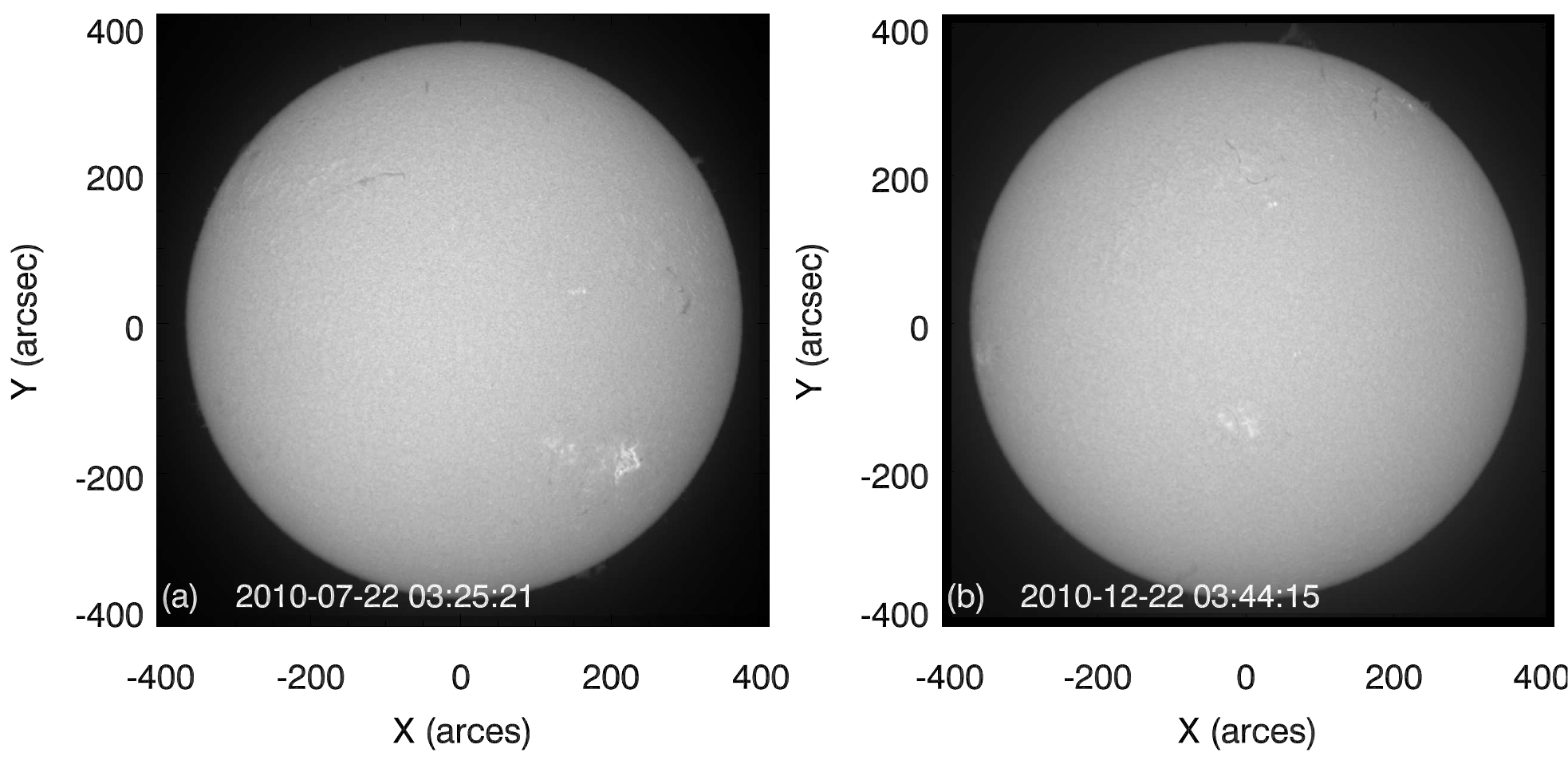}}
   \caption{(a): The larger field-of-view is cut for combining a 1.1R$_\sun$ field. (b): The small one filled for a 1.1R$_\sun$ field with artificial margin than can be seen more darker around.} \label{field}
\end{figure}
\begin{figure}
   \centerline{\includegraphics[width=1\textwidth,clip=]{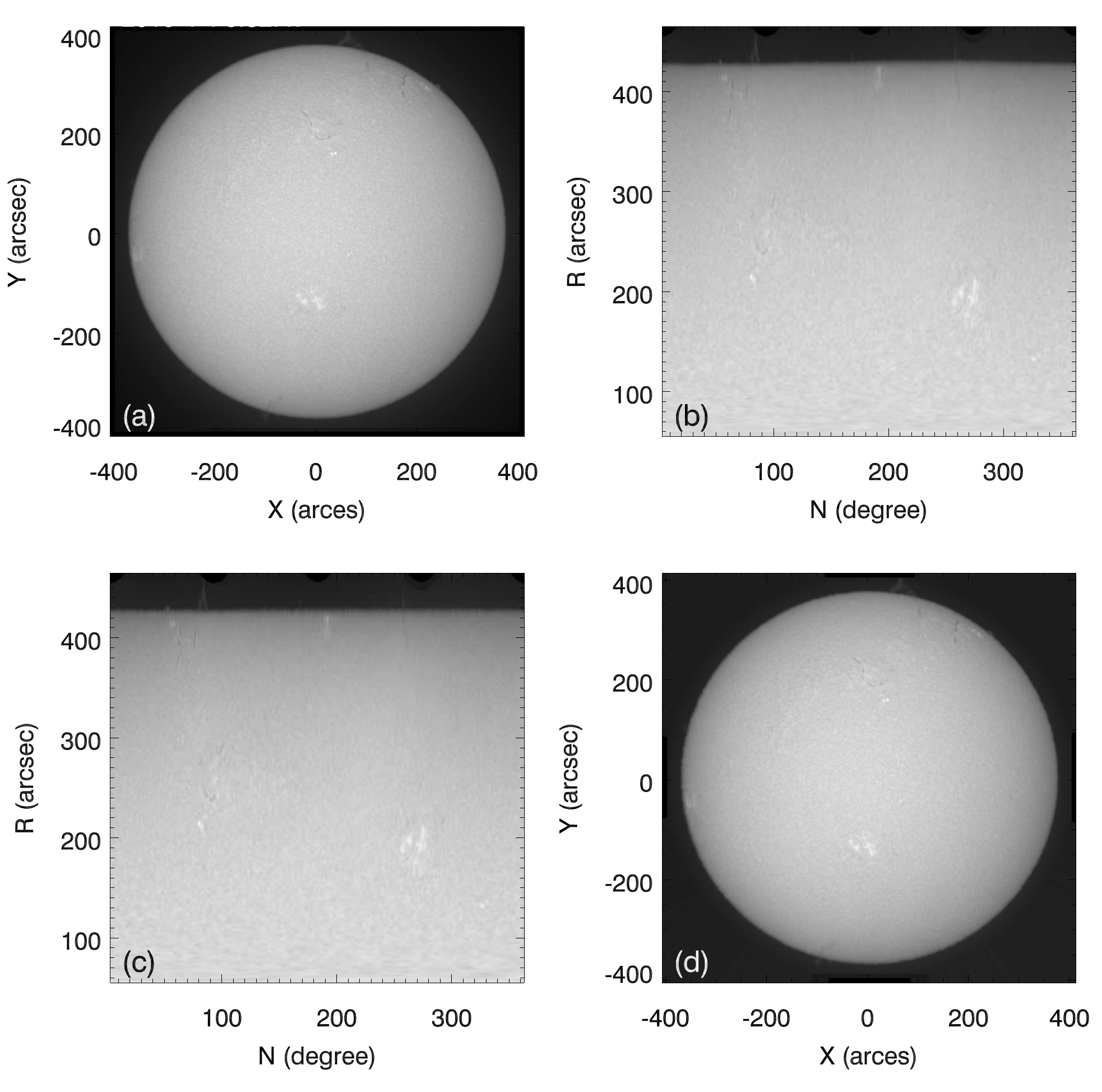}}
   \caption{(a): Image in cartesian coordinate system with central location and field of view correction. (b): Image with varied limb transformed from (a) in polar coordinate system. (c): Image with the limb of uniform length in polar coordinate system.(d): The correction image in cartesian coordinate system with  the uniform limb length}\label{circlere}
\end{figure}

After the position correction in Section \ref{sub-cenrad}, the suitable field-of-wiew is integrated. Here field-of-view with 1.1R$_\sun$ scale is integrated by refering higer-presion center and radius calculated by canny operator and circlefit.pro (R$_\sun$ is the radius of sun that is equal to the radius obtained by canny operator and circlefit.pro). For most parts of full-disk H$\alpha$ image, these were observed with feild-of-viwe larger than 1.1R$_\sun$, so the feild-of-viwe with 1.1R$_\sun$ is cut out from the whole image, however there are some image with feild-of-viwe smaller than 1.1R$_\sun$, then the margin should be filled all around the whole image observed. As is shows in Figure \ref{field}, where the left (a) means larger field-of-view is cut and give a 1.1R$_\sun$ field, and the right (b) corresponds the small one filled for a 1.1R$_\sun$ field (the filled margin can be seen more darker around). Then, the shape of circle is corrected based on the suitable field-of-wiew image, and the followings give this procedure: 1. The image in cartesian coordinate system is transformed to polar coordinate system, as is shown in panel (a) and (b) of Figure \ref{circlere}, it can be see that limb of disk varied as indicated in panel (b). 2. Basing on the initial multi-radii (here radius is obtained by center-of-gravity for individual 360$^{\circ}$ azimuth, radius is not the fitted single-radius with higer-presion by canny operator and circlefit.pro), the set standard length (500 pixels) of radius is elongated or compressed (here the length of Y-axis that is lower/higher than initial radius was fixed pixles [0-500]/[501-511], the total lenght of Y-axis is 512 pixels), then panel (c) image in polar coordinate system is obtained, which is transformed to standard circle image with uniform radii and dimensions of 1024 $\times$ 1024 in cartesian coordinate system as is shown in panel (d).

\begin{figure}
   \centerline{\includegraphics[width=1\textwidth,clip=]{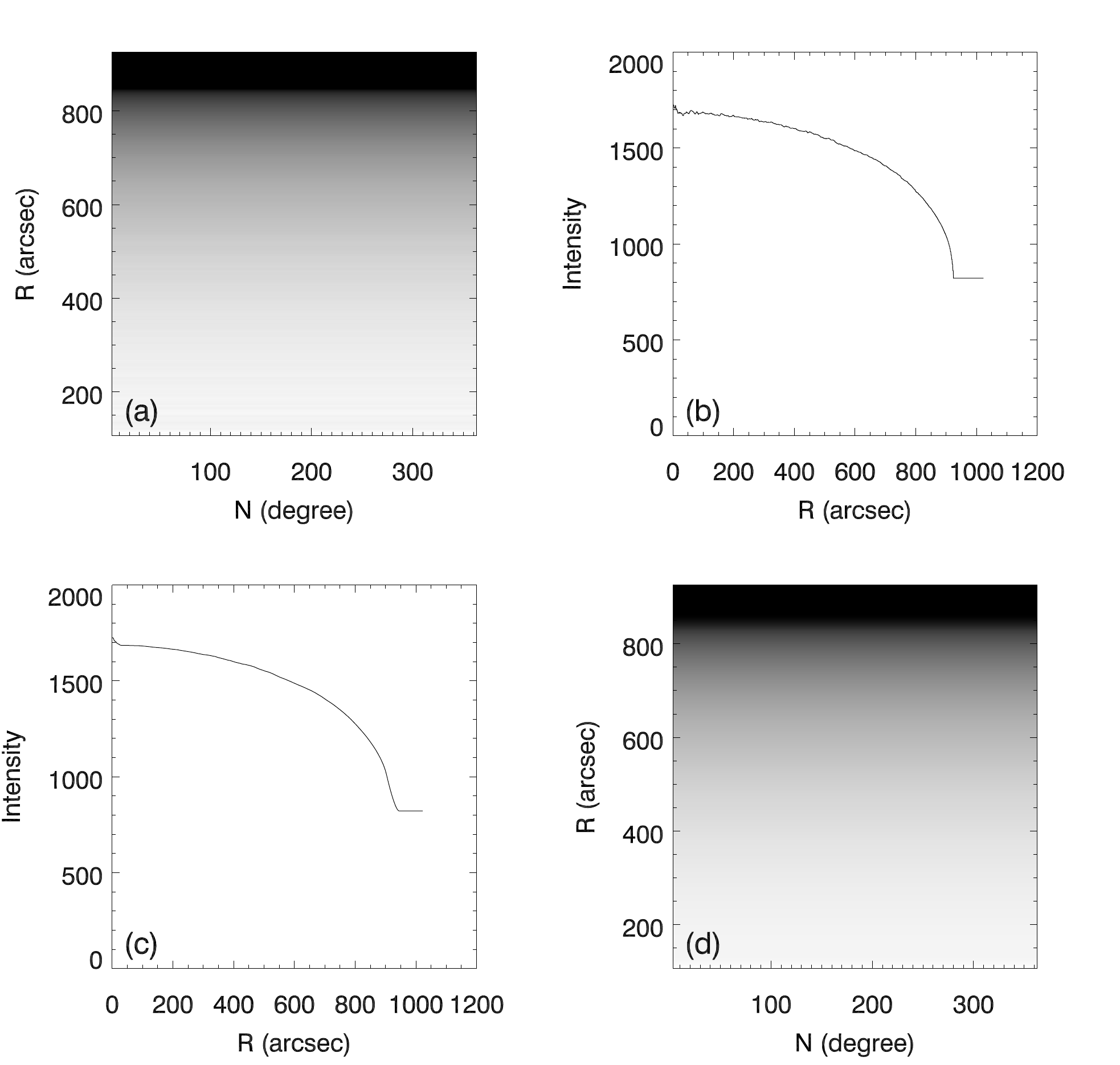}}
   \caption{(a): Intensity correction background image in polar coordinate system. (b): The original profile of limb dark obtained from one vertical line in panel (a) . (c): The procesed profile smoothed from its original one panel (a). (d): Revised intensity correction background image in polar coordinate system.}\label{intere_limbdarkprofile}
\end{figure}

\begin{figure}
   \centerline{\includegraphics[width=1\textwidth,clip=]{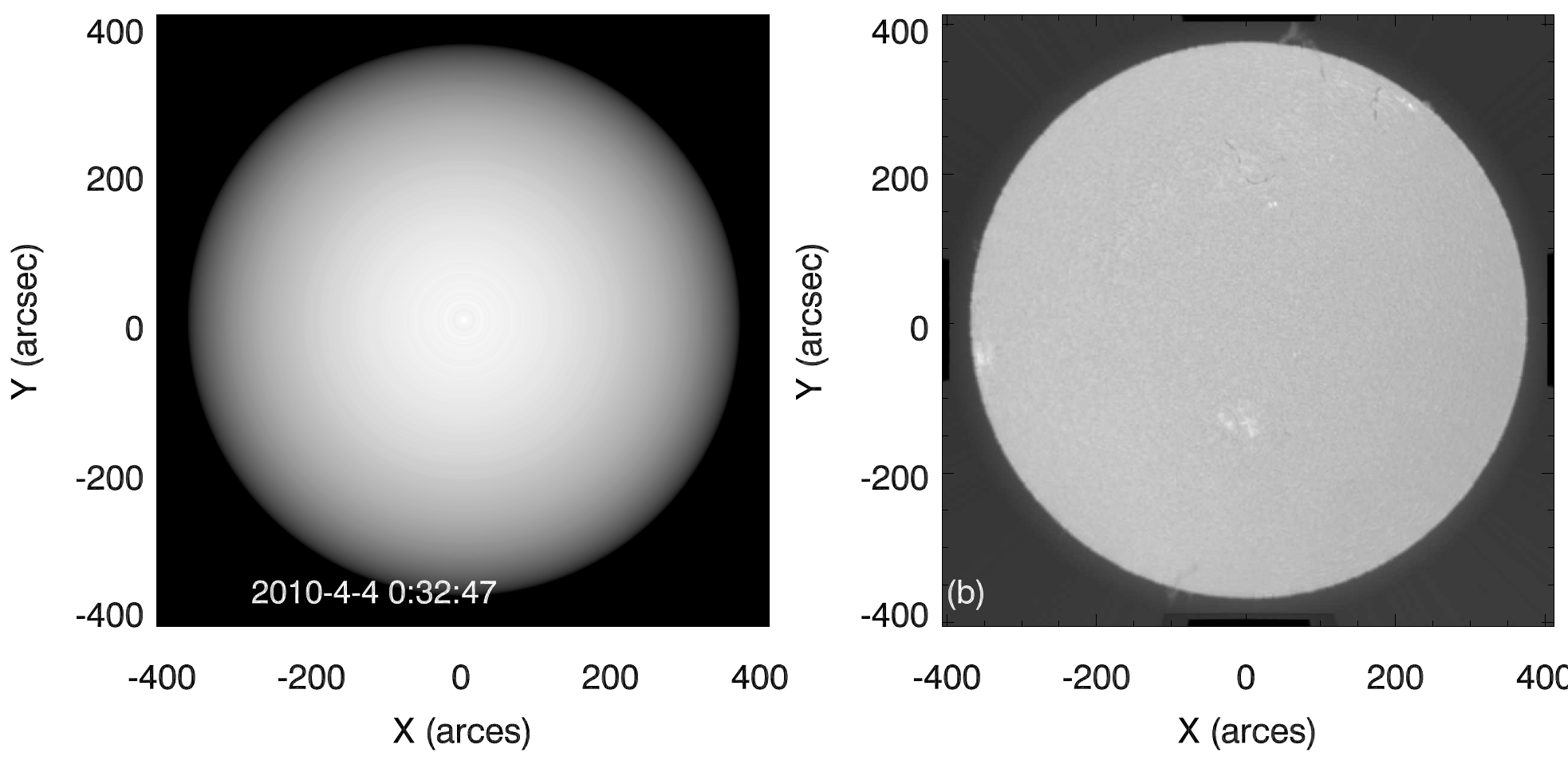}}
   \caption{(a):  Intensity correction background image transformed in cartesian coordinate system. (b): The intensities normalization image (image=imag1/imag2, image1 is that in (d) of Figure \ref{circlere}), image2 is the one in panel (a) of this Figure.}\label{intere}
\end{figure}
\begin{figure}
   \centerline{\includegraphics[width=0.6\textwidth,clip=]{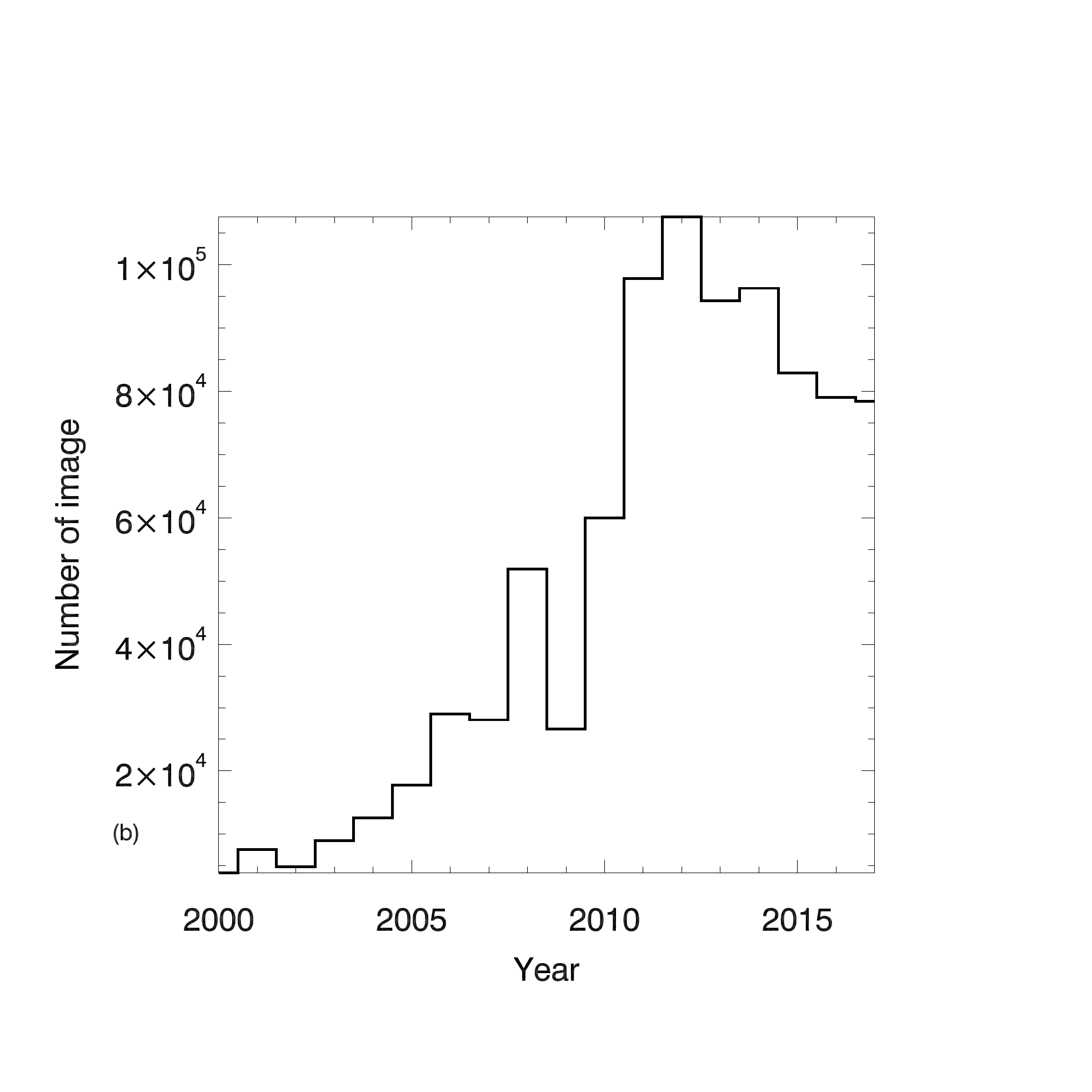}}
   \caption{The statistical number of processed H$\alpha$ full-disk images oberved each year by 14cm and 20cm telescopes at HSOS from 2000 to 2017.}\label{numofimages}
\end{figure}
\subsection{Intensities Normalization for full-disk image}
From the image of panel (d) of Figure \ref{circlere}, it can be seen that the intensity is not uniform for the solar disk, and the limb dark effects exist in this image. Thus, the median substitution method is applied to get the intensity correction background image. Firstly, based on image of panel (c) of Figure \ref{circlere} the median value of each row was first calculated and all pixel values in that row replaced with this value. Based on this process to produce an initial limb-darkening profile that is subsequently smoothed with a 16-pixel kernel median filter. At last, the intensity correction background image in polar coordinate system is obtained as shown in panel (a) of Figure \ref{intere_limbdarkprofile}, it should be noted that structure on the disk possible affect intensity correction background. So the classic approach is to fit or smooth the profile of limb dark. Panel (b) in Figure \ref{intere_limbdarkprofile} is the original profile of limb dark, which is obtained from one vertical line in panel (a) and plot its intensity  along radius, it can be seen that there are some jumping values in the curve. Panel (c) is the procesed profile smoothed from its original one (b) in Figure \ref{intere_limbdarkprofile}. While panel (d) shows the intensity correction background image revised from the procesed profile of limb dark. It can be seen that in intensity correction background image all structures with evident feature are eliminated naturally and the intensity along radius is very smooth. Secondly, the intensity correction background image in polar coordinate system is transformed to cartesian coordinate system, to get the intensity correction background image as is shown in panel (a) Figure \ref{intere}. At last, the intensities normalization image is obtained as given in panel (b) in Figure \ref{intere} , which is calculated by dividing (division operation) background image from image of panel (d) of Figure \ref{circlere}. Comparing with image in panel (d) of Figure \ref{circlere}, the radial ununiform and limb dark effects are evidently removed in the image in panel (b) of Figure \ref{intere}.
\section{Summaries}
In this paper, to standardize H$\alpha$ full-disk images oberved at HSOS some necessary procedures and algorithms are developed based on basicimage processing techniques.
Firstly, the solar disk is estimated and obtained by histogram analysis of the whole image. Then, the center-of-gravity and canny operation are used to calculate high-precision center and radius of solar disk, which is used to shift image and get the image with definitized field-of-view. Thirdly, the circle correction is carried out on these pre-process full-disk image. At last, the full-disk image intensities are normalized using  median substitution method, which assure to remove limb dark effects and other non-radial intensity uniforms.
The details about these procedures and algorithms are given in the paper and all historical data observde at HSOS have been prossecced using these algorithms. After the standardization of observation, it is certainly to facilitate researchers to use full-disk chromosphere observations for studying large-scale solar activities, such as filaments and solar flare and so on. Figure \ref{numofimages} shows the number of H$\alpha$ full-disk images oberved at HSOS, before 2006 the observations were carried out by 14cm telescope, while after 2006 it is 20cm telescope. Here it can be seen that the data sampling frequency of 20cm telescope own high temporal resolution of data acquisition, especially after 2009, in fact some time there are high frequency data with cadence one image per second that can be used to study especially the transient phenomena occurring in the solar chromosphere. At the web of $http://sun.bao.ac.cn/hsos_data/GHA/$, all the processed observations will be released soon with convenient query and download functions, and some examples are put at $http://sun.bao.ac.cn/hsos_data/download/HSOS_Ha/$.
\label{S-Conl}
\acknowledgments

This work was partly supported by National Natural Science Foundation of China (Grant No.U1531247, 2014FY120300) and the Strategic Priority Research Program
on Space Science, the Chinese Academy of Sciences (Grant No. XDA15320302, XDA15320102, XDA15052200), the 13th Five-year InformatizationPlan
of Chinese Academy of Sciences Grant No. XXH13505-04,the Young Researcher Grant of National Astronomical Observations,
Chinese Academy of Sciences, and the Key Laboratory of Solar Activity National Astronomical Observations, Chinese Academy of Sciences.

The authors declare that there is no conflict of interest regarding the publication of this paper.



\end{document}